\documentclass[
	aps, pra,
	superscriptaddress, twocolumn,
	floatfix,
	10pt
]{revtex4-2}

\usepackage{silence}
\usepackage{parskip}

\usepackage[final]{graphicx}
\graphicspath{{./figures/}}
\usepackage{times,bbm,amsmath,amssymb}
\usepackage{mathtools}
\usepackage{microtype}
\usepackage{epsfig,color}
\usepackage[dvipsnames]{xcolor}
\usepackage{hyperref}
\hypersetup{
    colorlinks = true
}
\usepackage{cleveref}

\usepackage{libertine}

\usepackage{float,siunitx}
\usepackage[caption = false]{subfig}

\usepackage{booktabs}

\usepackage{acronym}
\usepackage{easyReview}

\usepackage{tikz}

\usepackage{standalone}

\usepackage{physics}

\usepackage{blindtext}

\usepackage[toc,page]{appendix}

\newcommand{\bs}[1]{\boldsymbol{#1}}
\newcommand{\on}[1]{\operatorname{#1}}
\newcommand{\parTitle}[1]{\noindent{\emph{#1} ---}}

\newcommand{\bsn}{{\bs n}}

\newcommand{\CC}{\mathbb{C}}
\newcommand{\PP}{\mathbb{P}}
\newcommand{\RR}{\mathbb{R}}

\newcommand{\calC}{{\mathcal{C}}}

\newcommand{\calH}{{\mathcal{H}}}

\newcommand{\calO}{{\mathcal{O}}}

\newcommand{\calQ}{{\mathcal{Q}}}

\begin{document}
\title{Coherence-based operational nonclassicality criteria}

\author{Luca Innocenti}
\affiliation{Department of Optics, Palack{\'y} University, 17. Listopadu 12, 771 46 Olomouc, Czech Republic}
\affiliation{Università degli Studi di Palermo, Dipartimento di Fisica e Chimica – Emilio Segrè, via Archirafi 36, I-90123 Palermo, Italy}

\author{Lukáš Lachman}
\affiliation{Department of Optics, Palack{\'y} University, 17. Listopadu 12, 771 46 Olomouc, Czech Republic}

\author{Radim Filip}
\affiliation{Department of Optics, Palack{\'y} University, 17. Listopadu 12, 771 46 Olomouc, Czech Republic}

\begin{abstract}
The nonclassicality of quantum states is a fundamental resource for quantum technologies and quantum information tasks in general.
In particular, a pivotal aspect of quantum states lies in their coherence properties, encoded in the nondiagonal terms of their density matrix in the Fock-state bosonic basis.
We present operational criteria to detect the nonclassicality of individual quantum coherences that only use data obtainable in experimentally realistic scenarios.
We analyze and compare the robustness of the nonclassical coherence aspects when the states pass through lossy and noisy channels. The criteria can be immediately applied to experiments with light, atoms, solid-state system and mechanical oscillators, thus providing a toolbox allowing practical experiments to more easily detect the nonclassicality of generated states.
\end{abstract}
\maketitle

\section{Introduction}

The nonclassicality of quantum states is of utmost importance for quantum information tasks~\cite{yadin2018operational}, ranging from quantum communication and computation~\cite{knill2001scheme,kok2007linear,gisin2007quantum,shapiro2009quantum},  quantum sensing~\cite{degen2017quantum}, and thermodynamics~\cite{maslennikov2019quantum}.
Several notions of nonclassicality have been explored in different contexts.
For bosonic systems, the indivisibility of single bosons has for a long time been considered a direct experimental manifestation of nonclassicality~\cite{hanbury1956test,kimble1977photon,short1983observation,hong2017hanbury}. Another type of nonclassicality
is the impossibility of a state to be writable as a convex decomposition of coherent states~\cite{glauber1963coherent,mandel1995optical,vogel2006quantum}.
This can be formalised as the failure of a state $\rho$ to be decomposable as
\begin{equation}\label{eq:classical_states}
    \rho = \int d^2\alpha P(\alpha) \ketbra\alpha
\end{equation}
for some probability distribution $P$~\cite{glauber1963coherent,sudarshan1963equivalence}.
Operationally, coherent states $|\alpha\rangle$ are ideal states of a linear oscillator driven by external coherent force.
However, reconstructing the $P$ function experimentally is highly nontrivial~\cite{kiesel2008experimental,kiesel2011nonclassicality},
and criteria to detect $P$-nonclassicality include witness-based ones, relying on bounds on expectation values with respect to the $P$ function~\cite{richter2002nonclassicality,korbicz2005hilbert}; hierarchies of necessary and sufficient nonclassicality criteria based on the moments of distribution~\cite{agarwal1992nonclassical,shchukin2005nonclassical,shchukin2005nonclassicality,vogel2008nonclassical,miranowicz2010testing}; and criteria based on different approaches~\cite{rivas2009nonclassicality,bohmann2020phasespace,tan2020negativity,grunwald2019effective,grunwald2020nonquantum}.
The above methods share the shortcoming of relying on \textit{global} properties of the state, such as statistical moments, rather than being tailored to the specific information acquired in a given experimental scenario.
Other nonclassicality criteria, based on photon-click statistics~\cite{sperling2012subbinomial,sperling2017detectorindependent,filip2013hierarchy,rigovacca2016nonclassicality,lachman2019criteria,lachman2019faithful}, are based on operationally measurable quantities, but are tied to specific detection schemes.

As of yet, no nonclassicality criterion specifically tailored at individual quantum coherences --- as opposed to requiring a more complete (often tomographically complete) knowledge of the state --- is known. A possible reason for this is that while the shape of the set of classical states when only diagonal matrix elements are being observed is relatively manageable via generalised Klyshko-like inequalities~\cite{klyshko_observable_1996,innocenti2020nonclassicality}, finding similar inequalities when also coherences are involved is highly nontrivial.
However, being quantum coherences a useful resource for a variety of quantum information tasks~\cite{streltsov2017colloquium}, understanding the nonclassicality involving individual coherences would be a valuable from both experimental and fundamental viewpoints.
In this Letter we lay out a framework to characterise the nonclassicality with Fock-state quantum coherences, by devising operational criteria to certify the nonclassicality of states leveraging their coherences.
We can thus discuss the role of coherence-based observables on certifying incompatibility with classical states of the form~(\ref{eq:classical_states})
Opposite to what was the case when characterising nonclassicality using only Fock state probabilities~\cite{innocenti2020nonclassicality}, we find that when coherences are involved it is also pivotal to consider the boundary of the set of all states in the considered spaces, as in some situations the two can partially overlap, resulting in more care being needed when devising nonclassicality criteria.
To ensure seamless applicability to experimental scenarios, our criteria only exploit knowledge of the expectation values of few observables, as one would have access to in realistic cirumstances.
To achieve this, we devise an approach to nonclassicality detection based on incomplete knowledge of the density matrix~\cite{filip_detecting_2011,innocenti2020nonclassicality}, extending the current state of the art by analysing the information hidden in off-diagonal terms.
These elements are directly measurable by Ramsey-like interferometry of trapped ion~\cite{mccormick2019quantum}, superconducting circuit experiments~\cite{hu2019quantum}, and electromechanical oscillators~\cite{chu2018creation}. For light, atomic ensembles, and optomechanical oscillators, they can be reconstructed using homodyne tomography.
We compare our criteria to those relying only on Fock-state probabilities~\cite{klyshko_observable_1996,innocenti2020nonclassicality}, and analyze the nonclassical depth of various quantum coherences represented by different off-diagonal elements.

We find that observing coherence terms can provide enhanced predictive power in terms of nonclassicality detection, and showcase this in several instances of nonclassicality in one-, two-, and three-dimensional spaces.
More precisely, we find that, remarkably, in some situations the Fock state probabilities alone are sufficient to detect all of the existing nonclassicality, whereas in other situations adding knowledge about coherence terms provides enhanced predictive power.
Moreover, we show how each set of different measured observables provides a distinct boundary of nonclassicality, and study the behaviour in these spaces of superposition states subject to attenuation and thermal noise. This further highlights how different types of noise affect the observable nonclassicality in nontrivial ways, even in relatively low-dimensional spaces.

\section{General framework}
\label{sec:background_on_convex_geom}

\emph{Support function and support hyperplanes ---}
Suppose we are given the expectation values $\langle\calO_i\rangle$
for some set of observables $\calO_i$, and want to figure out whether these measurements are compatible with \textit{some} classical state.
Given the relevant Hilbert space $\calH$, we will denote with $\calQ$ the set of density matrices in this space, and with $\calC\subset\calQ$ the convex hull of the coherent states.
Let us also denote with $\bs\calO(\rho)\equiv(\Tr(\calO_k\rho))_{k=1}^n$ the set of expectation values resulting from measuring $\rho$.
We seek a method to determine whether, given an unknown state $\rho$, whether there is some $\sigma\in\calC$ compatible with the observed measurements, that is, to determine whether
    $\bs\calO(\rho) \in \{ \bs\calO(\sigma) : \, \sigma\in\calC\}$.

The convexity of $\calC$ and $\calQ$ allows to characterize them via supporting hyperplanes, using the tools of convex geometry~\cite{hug2010course}.
Any closed convex set $A\subset\RR^n$ is characterized be its \textit{support function} $h_A:\RR^n\to\RR$, defined as
    $h_A(\bs n) = \sup_{\bs x\in A} \langle \bs n,\bs x\rangle$.
Geometrically, $h_A(\bs n)$ represents the distance from the origin to the hyperplane tangent to $A$ orthogonal to $\bs n$.
Denote with $h_\calC(\bs n)$ and $h_{\calQ}(\bs n)$ the support functions of $\calC$ and $\calQ$, respectively, in the space of interest.
More explicitly, we consider the structure of $\calC$ and $\calQ$ when projected onto the finite-dimensional subspaces spanned by the observables measured in a given context.
This allows to devise criteria with a direct operational significance.
We can then translate the task of nonclassicality detection into finding whether there is $\bs n$ such that $h_{\calQ}(\bs n) > h_{\calC}(\bs n)$.
Whenever this is the case, it is possible to find a set of measurement results $\bs\calO\in\mathbb{R}^n$ such that $\bsn\cdot\bs\calO>h_{\calC}(\bsn)$, which certifies that these measurement results are not compatible with any classical state.
By studying the structure of $h_{\calC}(\bsn)$ and $h_{\calQ}(\bsn)$ for all $\bsn$, we can fully characterize the geometry of the classical set, and often end up with Klyshko-like nonclassicality criteria~\cite{klyshko_observable_1996,innocenti2020nonclassicality}.
Notably, in many of the scenarios considered here, we will be able to derive the relevant criteria without explicitly involving the corresponding support function.
This is possible in sufficiently simple situations where we can devise ad-hoc procedures to reach the conclusion.
These criteria are equivalent to the full set of criteria of the form $\bsn\cdot\bs\calO>h_\calC(\bsn)$, for all $\bsn$.
In a sense, we can understand these ad-hoc derivations as corresponding to a full characterization of the support functions $h_\calC(\bsn)$ for all values of $\bsn$.
Directly using the support function remains nonetheless very useful, as we will show in some explicit cases.

While $h_\calQ(\bs n)$ is generally easier, as it amounts to finding the largest eigenvalue of $\bs n\cdot\bs\calO\equiv \sum_i n_i\calO_i$, that is, computing the operator norm $\|\bs n\cdot\bs \calO\|_{\rm op}$.
On the other hand, computing $h_\calC(\bs n)$ is in general more difficult, as it involves maximising $\sum_i n_i\Tr(\calO_i\rho)$ over the set of $\rho\in\calC$.
Nonetheless, even though characterising $h_{\calQ}(\bs n)$ is relatively straightforward for any fixed value of $\bs n$, this does not trivially translate into an algebraic characterisation of the boundary of $\calQ$ itself.
We show here how to tackle this task in several cases of interest.

\emph{Coherence terms ---}
To focus on the nonclassicality of coherences, we will consider as basic observables $X_{jk}\equiv\ketbra{j}{k}+\ketbra{k}{j}$ and $Y_{jk}\equiv i(\ketbra{k}{j} - \ketbra{j}{k})$, which are a straightforward generalisation of non-diagonal Pauli matrices in higher dimensions. These naturally capture information hidden in coherence terms, that is not directly accessible via projections of the form  $P_j\equiv\ketbra j$.
The expectation value of $X_{jk},Y_{jk}$ on a coherent state $\ket\alpha$ with $\alpha=\sqrt{\mu} e^{i\phi}$
are related to the Fock-state number probabilities
$P_i\equiv\ketbra i$ as
\begin{equation}\label{eq:XijYijdefinition}
\scalebox{0.88}{$\displaystyle\begin{aligned}
    X_{ij} = 2\sqrt{P_i P_j} \cos(\phi(i-j)), \,\,
    Y_{ij} = 2\sqrt{P_i P_j} \sin(\phi(i-j)).
\end{aligned}$}
\end{equation}
For ease of notation, here and in the rest of the paper, we will with some abuse of notation conflate the operators $X_{jk}$ with their expectation values on a given state $\rho$, $\langle X_{jk}\rangle_\rho\equiv\Tr(X_{jk}\rho)$. For example,~\cref{eq:XijYijdefinition} would be more precisely written as
    $\langle X_{jk}\rangle_\alpha
    = 2\sqrt{\langle P_j\rangle_\mu\langle P_k\rangle_\mu} \cos(\phi(j-k)).$ 
More generally, we can consider the rotated operators $R_{jk}(\theta)\equiv \cos(\theta) X_{jk} + \sin(\theta) Y_{jk}$, whose expectation value on coherent states reads
    $R_{jk}(\theta) =
    2\sqrt{P_j P_k}
    \cos(\theta - \phi(j-k))$.

\emph{Quantum boundary ---}
When only dealing with Fock-state probabilities, any probability distributions is compatible with some quantum state, and thus the boundary of $\calQ$ is simply defined by the relations $\sum_j P_j\le1$ and $0\le P_j\le 1$.
The situation changes significantly when coherence terms are being considered.
Finding the boundary of $\calQ$ then amounts to figuring out the conditions under which the observed expectation values fit into a positive semidefinite matrix.
Further details on how this process results in different inequalities are given on a per-case basis in the text, and we also include for completeness a more general discussion in the SM.

\section{Nonclassicality criteria}

We will discuss here the nonclassicality certifiable via non-diagonal elements of the density matrix in the Fock state basis, as well as the nonclassicality encoded in nontrivial combinations of different coherence terms, or in nontrivial combinations of both coherence terms and Fock-state probabilities.

\textit{One-dimensional criteria ---}
We first study the class of nonclassicality criteria associated to an individual coherence term $R_{jk}(\theta)$. These are the easiest to apply in any experimental scenario where coherences are measured. In these spaces, the set of all states is bounded by $|R_{jk}(\theta)|\le1$, with bound saturated by the state $\frac1{\sqrt2}(\ket j+e^{i\theta}\ket k)$.
On the other hand, the corresponding classical bound is
\begin{equation}\label{eq:classical_bound_Rjktheta}
    |R_{jk}(\theta)| \le 
    \max_{\rho\in\mathcal{C}} 2\sqrt{\rho_{jj} \rho_{kk}}
    = 2 e^{-\frac{j+k}{2}}
    \frac{[(j+k)/2]^{\frac{j+k}{2}}}{\sqrt{j!k!}},
\end{equation}
where the maximisation can be restricted to the set of coherent states.
The corresponding bound for the set of \textit{all} states is instead
$|R_{jk}(\theta)| \le 1$, saturated by the state $\frac{1}{\sqrt2}(\ket j+e^{i\theta}\ket k)$.
In particular, we have
\begin{equation}\label{eq:boundary_single_coherences}\scalebox{0.96}{$\displaystyle
\begin{gathered}
    |X_{01}| \le \sqrt 2e^{-1/2} \approx 0.86,
    \,\,
    |X_{02} | \le \sqrt 2e^{-1} \approx 0.52, \\
    | X_{12} | \le \frac{\sqrt{27}}{2} e^{-3/2} \approx 0.58.
\end{gathered}
$}\end{equation}
This means that \textit{e.g.} measuring any value for the coherence of $0.87 \le \lvert X_{01}\rvert \le 1$ is sufficient to certify nonclassicality.
We refer to states violating any such inequality involving a single coherence term as displaying \textit{nonclassical coherence}.
As an interesting example showcasing the nontriviality of nonclassical coherences, consider any $\rho$ that is a mixture of the nonclassical state $|1\rangle$ with any classical $\rho_{\rm cl}$. Although such $\rho$ might be recognizable as nonclassical via some observable, because only $\rho_{\rm cl}$ produces coherence terms, it will not display any \textit{nonclassical coherences}, meaning it is not recognizable as nonclassical by any criterion involving only coherence terms. 

\textit{Two-dimensional criteria ---}
Even though one-dimensional criteria using individual coherences can always be applied, it is possible to devise stronger criteria by characterizing the boundary corresponding to higher-dimensional spaces.
This allows to detect as nonclassical states whose nonclassicality cannot be deduced from any individual coherence term.
For example, when a pair of coherences of the form $(X_{jk},Y_{jk})$ is considered, $\calQ$ is characterised by the inequality
$X_{jk}^2+Y_{jk}^2\le 1$, saturated by states of the form $\frac1{\sqrt2}(\ket j+e^{i\theta} \ket k)$ for all $\theta\in[0,2\pi]$.
On the other hand, $\calC$ only forms a circle of radius given by~\cref{eq:classical_bound_Rjktheta}.
It follows that measuring any pair of values for $(X_{jk},Y_{jk})$ that falls outside such circle, is sufficient to certify the nonclassicality of the underlying state.

One can also ask what nonclassicality is encoded in pairs of observables including both coherences and Fock probabilities.
Consider for example some $R_{jk}(\theta)$ together with $P_j$.
The boundary of $\calQ$ corresponds to
\begin{equation}
    |R_{jk}(\theta)| \le \max_{P_k}
    2\sqrt{P_j P_k}
    = 2\sqrt{P_j(1-P_j)}.
\end{equation}
where the maximum is taken with respect to all non-negative reals $P_k$ such that $P_j+P_k\le1$ (that is, over all quantum states compatible with the given values of $P_j$ and $R_{jk}(\theta)$).
On the other hand, classical states provide a generally more complex boundary.
For example, in the space $(P_0, X_{01})$, the boundary is defined by the inequalities $0\le P_0 \le 1$ and
\begin{equation}\label{eq:boundary_on_x01_given_p0}
    0 \le |X_{01}| \le 2 P_0 \sqrt{-\log P_0},
\end{equation}
with the latter saturated by coherent states.
\Cref{eq:boundary_on_x01_given_p0} means that there are states whose nonclassicality cannot be detected measuring $P_0$, nor measuring $X_{01}$ and applying~\cref{eq:boundary_single_coherences}, but is nonetheless revealed properly exploiting the knowledge of both $P_0$ and $X_{01}$.
An explicit example of this is measuring $P_0=0.2$ and $X_{01}=0.6$, where we do not detect nonclassicality using only $P_0$ or $X_{01}$, but we do via criterion in the two-dimensional space $(P_0,X_{01})$.

In~\cref{fig:convex_hulls_X01vs} we show how the nonclassicality criteria look like in two-dimensional spaces involving coherence terms, more specifically in the spaces $(P_0, X_{01}), (P_1,X_{01}), (X_{02},X_{01})$, and $(X_{12},X_{01})$.
To show the relation between two-dimensional and one-dimensional criteria, the figure also highlights the nonclassicality threshold corresponding to the one-dimensional criterion using only $X_{01}$. This corresponds to projecting each of the given plots onto the horizontal axis.
We refer to the SM for further details on the derivation of these criteria, as well as some discussion on nonclassicality criteria in the spaces $(P_0, P_2)$ and $(P_0, X_{02})$.
\Cref{fig:convex_hulls_X01vs} shows that measuring coherence terms provides valuable information regarding the nonclassicality of the states. This is reflected in the blue solid line in the figures being a strict subset of the gray solid line.
From an experimental point of view, these mean that measuring any pair of expectation values to be outside of the blue region is sufficient to certify the nonclassicality of the underlying state.
This makes for criteria that can be directly employed in pratical scenarios.

\textit{Three-dimensional criteria ---}
Another interesting case is obtained considering both Fock-state probabilities and coherences.
For example, in the space $(P_0,P_1,X_{01},Y_{01})$, the set $\calQ$ is characterised by the trivial constraint $0\le P_0+P_1\le1$ on the probabilities, with the additional constraint
\begin{equation}
    X_{01}^2 + Y_{01}^2 \le 4 P_0 P_1
    \iff |R_{01}(\theta)|\le 2\sqrt{P_0 P_1}\,\,\forall\theta,
\end{equation}
Thus in this space, $\calQ$ is a disk with radius $2\sqrt{P_0 P_1}$, as also shown in~\cref{fig:P0P1X01withCoherents2}.
More rigorously, as discussed in the SM, this constraint follows from Sylvester's criterion for positive semidefiniteness~\cite{johnson1985matrix}.
%
We thus find that, remarkably, albeit it is possible to detect a state as nonclassical using the values of $P_0$ and $P_1$, if using these values it is \textit{not} possible to detect the state as nonclassical, then it will still not be possible to do so adding knowledge about $X_{01}$ and $Y_{01}$.
With such analysis we are thus able to predict which observables will be useful for the purpose of identifying nonclassicality, which helps to devise more efficient experimental platforms.

In other cases, for example when one knows $(P_0,P_2)$ but not $P_1$, the associated coherence terms do provide information about nonclassicality.
This can be traced down to the boundary of $\calC$ in the space $(P_0,P_2)$ containing non-pure states, and to the non-convexity of the set of coherent states in the same space. 
Remarkably, in this case adding knowledge of $R_{02}$ allows to recognize as nonclassical states in the nonconvex region of the space $(P_0,P_2)$. Further discussion of this aspect is provided in the SM.

These case studies highlight the stimulating nontrivial features of nonclassical coherences that set them apart compared to previously known nonclassicality criteria~\cite{klyshko_observable_1996,innocenti2020nonclassicality}.
\section{Robustness of nonclassical coherences}

To probe the robustness of the devised criteria, we study how different degrees of attenuation and thermal noise affect our capacity to detect the nonclassicality of coherent states.
In the process, we find a rich landscape of possible behaviours.

\begin{figure}
    \centering
    \hspace{-5pt}\includegraphics[width=1\linewidth]{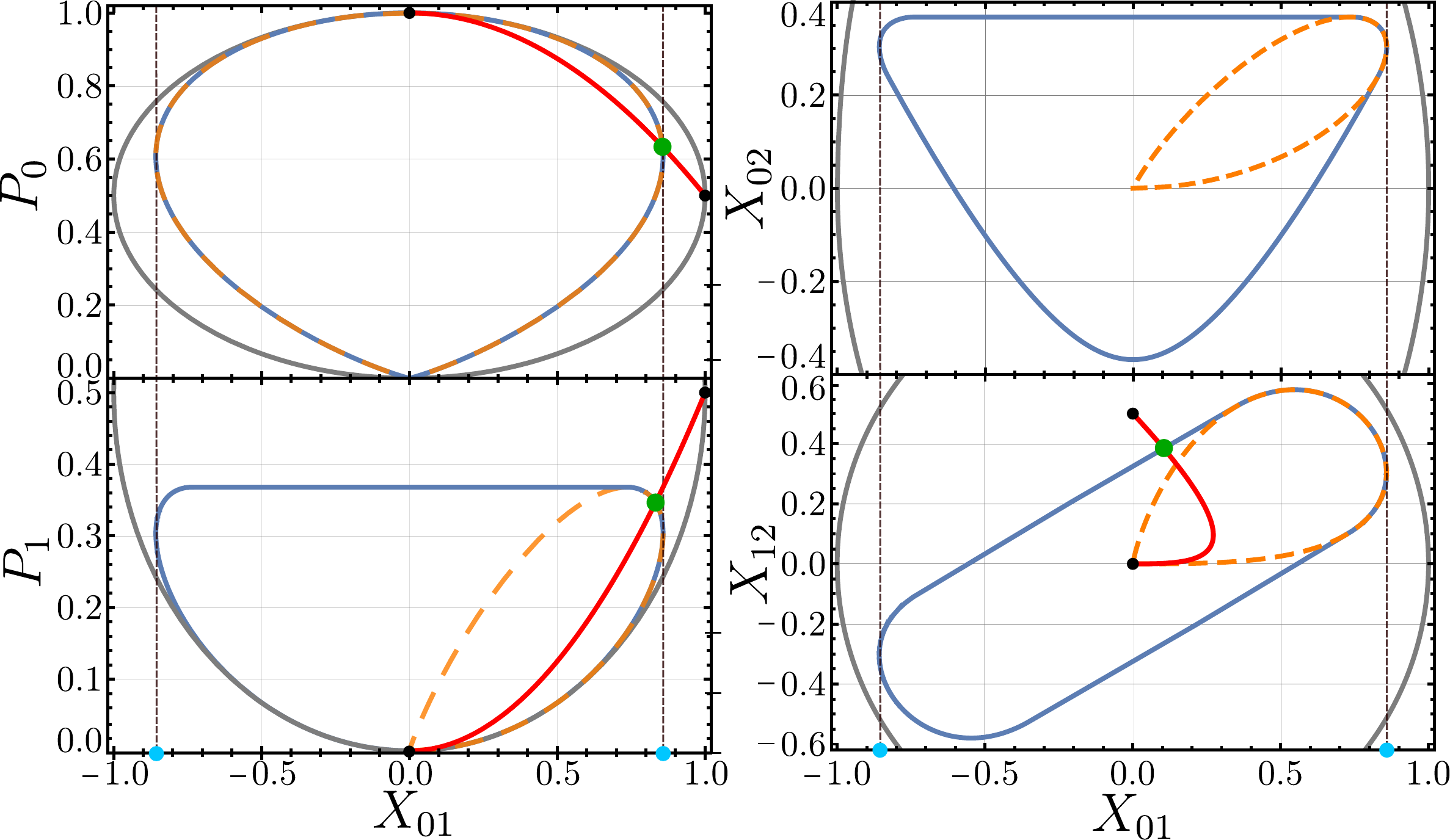}
    \caption{
        Classical regions in the spaces $(X_{01},P_0)$, $(X_{01},P_1)$, $(X_{01},X_{02})$, $(X_{01},X_{12})$.
        Solid blue lines trace the boundary of $\calC$. Dashed orange lines the set of \textit{coherent} states of the form $\ket{\sqrt\mu}$ for $\mu\in\mathbb{R}$.
        outer gray lines trace the boundary of $\calQ$.
        In the spaces $(X_{01},P_i)$, the solid red line corresponds to $\frac{1}{\sqrt2}(\ket0+\ket1)$ attenuated through a beamsplitter with transmissivity $T=|t|^2$, for $T\in[0,1]$. The two black dots joined by this line correspond to the states $\ket+$ and $\ket0$.
        The green dot marks the transmissivity corresponding to a transition between classicality and nonclassicality, and corresponds to $T\approx0.73$ in $(X_{01},P_0)$ and $T\approx 0.69$ in $(X_{01},P_1)$, respectively.
        We also show the results of attenuating $\frac{1}{\sqrt2}(\ket1+\ket2)$ in the $(X_{01},X_{12})$ space.
        The nonclassicality threshold, again marked with a green dot, corresponds to $T\approx0.84$.
        The cyan dots at the bottom of the downmost figure correspond to the one-dimensional nonclassicality threshold for the coherence term $X_{01}$, corresponding to $X_{01}\approx\pm0.86$.
    }
    \label{fig:convex_hulls_X01vs}
\end{figure}
\begin{figure}[tb]
    \centering
    \includegraphics[width=0.8\linewidth]{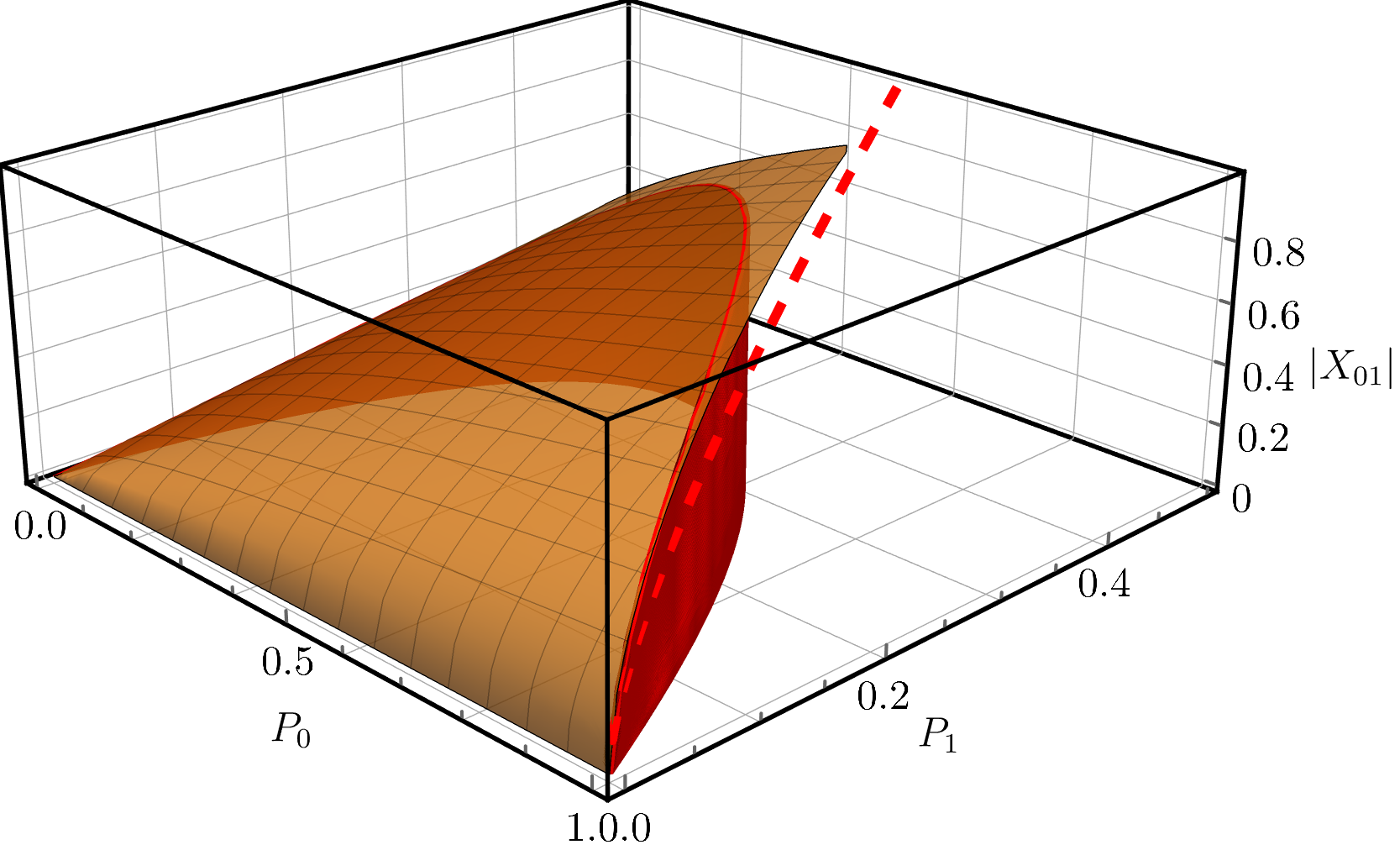}
    \caption{
        Boundary of the classical subset in the $(P_0, P_1, X_{01})$ subspace, where $P_0$ and $P_1$ are the probabilities of finding zero and one bosons, and $X_{01}$ is the expectation value of the coherence term.
        The dashed red line represents the attenuated states obtained from $\frac{1}{\sqrt2}(\ket0+\ket1)$, with different degrees of attenuations.
        The two orange surfaces represent the classicality boundaries corresponding to the criteria. In particular, the vertical surface corresponds to criterion in the $(P_0,P_1)$ subspace, while the other surface is the one bounding the value of $|X_{01}|$ for each value of $(P_0,P_1)$.
        In this space, nonclassicality is thus certified by checking that a point lies beyond at least one of these two surfaces.
    }
    \label{fig:P0P1X01withCoherents2}
\end{figure}

We show in~\cref{fig:convex_hulls_X01vs} how attenuated superposition states look like in these two-dimensional subspaces.
In~\cref{fig:convex_hulls_X01vs}(\textbf{a}) and (\textbf{b}) we plot the points corresponding to attenuation of $\frac{1}{\sqrt2}(\ket0+\ket1)$ with transmittivities $T\in[0,1]$. We find that nonclassicality can be certified using $(P_0,X_{01})$ for $T>0.73$, while in $(P_1, X_{01})$ for $T>0.69$.
In~\cref{fig:convex_hulls_X01vs}(\textbf{d}) we plot attenuated states obtained from $\frac{1}{\sqrt2}(\ket1+\ket2)$. In this case we find the nonclassicality threshold in the space $(X_{01},X_{12})$ to sit at around $T\approx 0.84$.

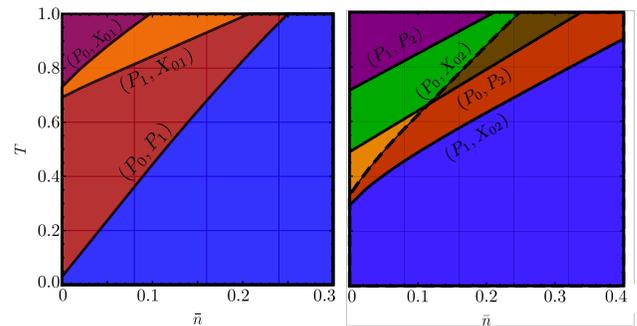
\begin{figure}
    \centering
    \begin{tikzpicture}
    \node[anchor=south west] (A) at (0, 0)%
		{\includegraphics[width=0.52\linewidth]{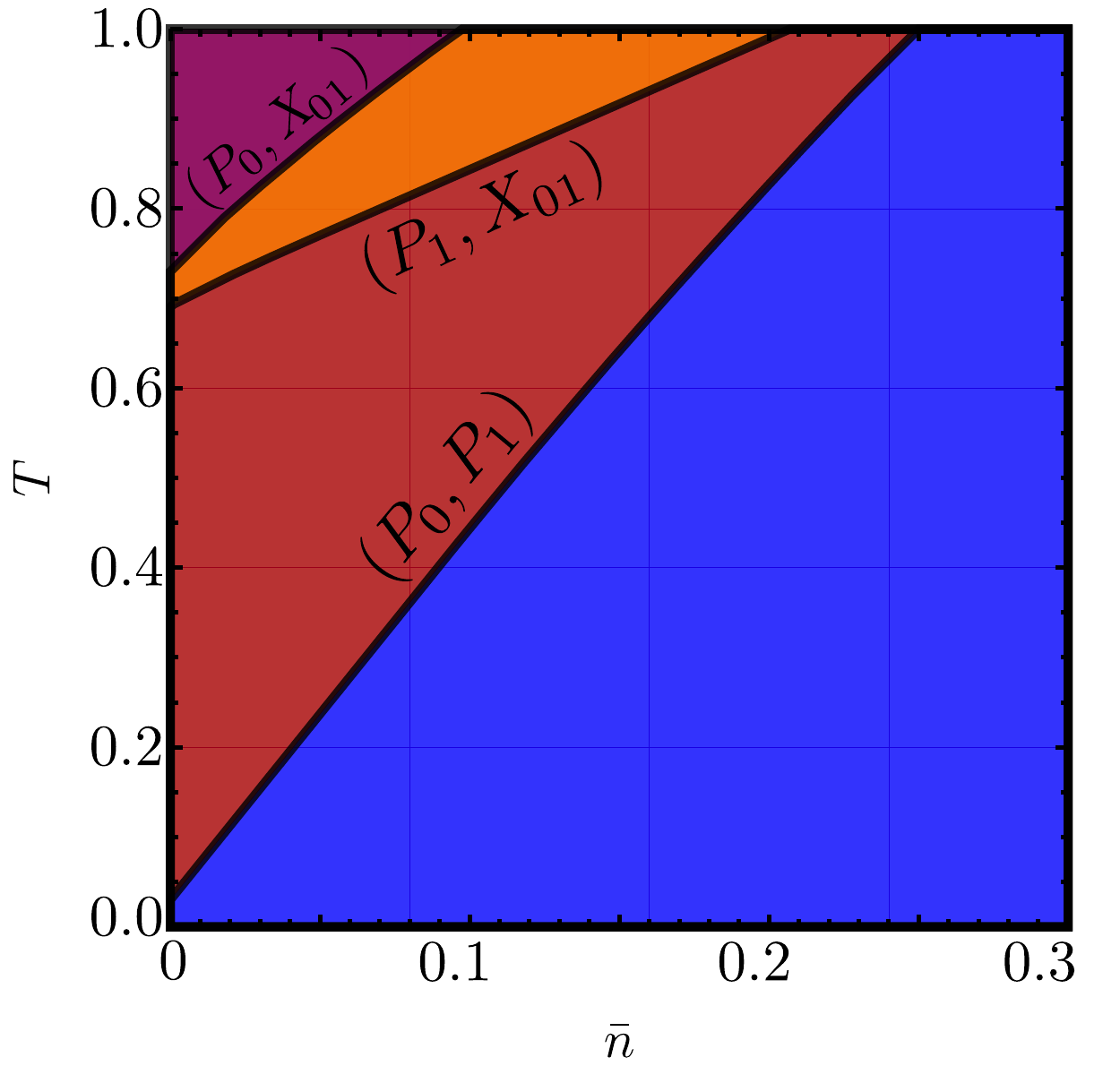}};
    \end{tikzpicture}\hspace{-27pt}
    \begin{tikzpicture}
    \node[anchor=south west] (A) at (0, 0)%
		{\includegraphics[width=0.515\linewidth]{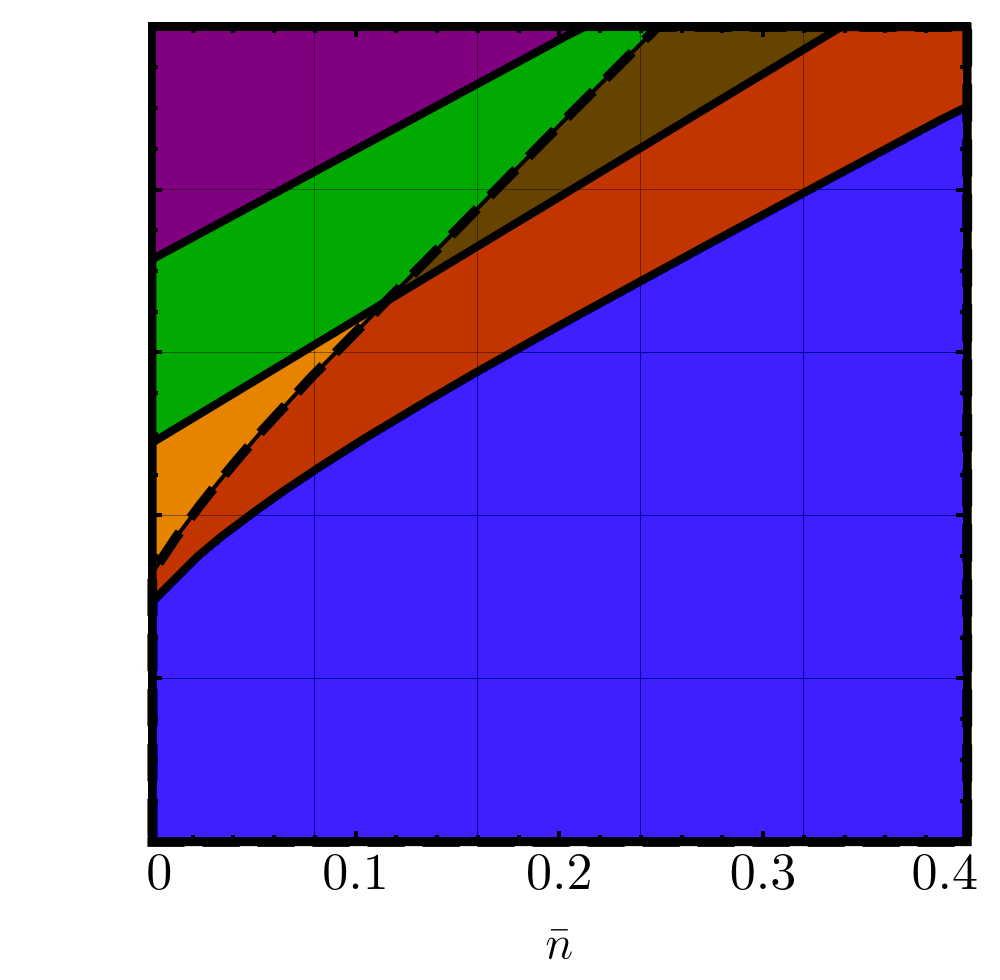}};
		\node[above, rotate=30.5] at (1.5, 3.7) {\scalebox{0.74}{$(P_1, P_2)$}};
		\node[above, rotate=33] at (2.7, 3.15) {\scalebox{0.74}{$(P_0, P_2)$}};
		\node[above, rotate=47] at (2.2, 3.42) {\scalebox{0.74}{$(P_0, X_{02})$}};
		\node[above, rotate=33] at (2.6, 2.5) {\scalebox{0.74}{$(P_1, X_{02})$}};
    \end{tikzpicture}
    \caption{Nonclassicality of the states $\frac{1}{\sqrt2}(\ket0+\ket1)$ (left) and $\frac1{\sqrt2}(\ket0+\ket2)$ (right) after attenuation with transmissivity $T\equiv |t|^2$ and thermalisation with average boson 
    number $\bar n$, as detected different criteria.
    Each line corresponds to a nonclassicality criterion using two observables, separating the lower-right region of nonclassicality from the rest.
    }
    \label{fig:nonclassicality_0plus1_thermAndAttenuated_nvsT_mondrian}
\end{figure}

We furthermore analyze the robustness of states obtained from $\frac1{\sqrt2}(\ket0+\ket1)$ and $\frac1{\sqrt2}(\ket0+\ket2)$ when subject to both attenuation and thermal noise. As shown in~\cref{fig:nonclassicality_0plus1_thermAndAttenuated_nvsT_mondrian}, different pairs of observables result in different nontrivial nonclassicality criteria, highlighting how the hardness in witnessing nonclassicality strongly depends on the measured observables.

We find a rich landscape of possibilities, that highlights how choosing suitable observables can strongly enhance the capacity to detect the nonclassicality of a given state by leveraging their coherence properties.

\section{Conclusions}

We showed how to leverage individual off-diagonal elements of the density matrix in the Fock-state basis to assess the nonclassicality of quantum states, and how to pair these with Fock-state probabilities to detect even larger classes of nonclassical states.
This paves the way for a more thorough understanding of the relation between nonclassicality in quantum mechanics and the widely studied resource theories of quantum coherence~\cite{streltsov2017colloquium}.

These criteria can be directly implemented experimentally, by simply measuring the relevant observables and checking whether the obtained expectation values violate the given criteria. Such a protocol can be implemented with state of the art technology, for example in photonics~\cite{yukawa2013generating}, trapped ions~\cite{mccormick2019quantum}, superconducting~\cite{hu2019quantum}, and electromechanical~\cite{chu2018creation} platforms.
Enhanced capabilities of detecting nonclassical states have several applications, for example in the context of quantum metrology~\cite{kwon2019nonclassicality,mccormick2019quantum,wang2019heisenberglimited} and quantum error-correction for quantum communication and computing~\cite{li2017cat,michael2016new}.
Our results highlight the nontrivial way the nonclassicality of states translates into nonclassical coherences: while measuring coherences provides useful information in many situations, there also exist scenarios where all the information about nonclassicality is already encoded in the Fock-state probabilities. This leaves open the stimulating question of characterizing the precise class of situations where coherent terms do or do not result in additional predictive power.

Our work paves the way for a more thorough understanding of nonclassicality detection with multimode coherences over diverse platforms~\cite{babichev2004homodyne,thekkadath2020quantum,gao2018programmable,zhang2018noon,gao2019entanglement,gan2020hybrid}. Such criteria would provide enhanced detection schemes for platforms generating entangled states, imposing more lenient demands on those sources than entanglement.
Another natural vanue of further study is to devising criteria for quantum non-Gaussianity~\cite{straka2018quantum,lachman2019faithful}, and study the role of coherences in that context, which remain not fully understood.



\section*{Acknowledgments}
L.I. acknowledges support from MUR and AWS under project 
PON Ricerca e Innovazione 2014-2020, ``calcolo quantistico in dispositivi quantistici rumorosi nel regime di scala intermedia" (NISQ - Noisy, Intermediate-Scale Quantum).
L.L. and R.F. acknowledge the project 21-13265X of Czech Science Foundation.

\clearpage
\newpage

\begin{appendices}

\section{Derivation of boundaries for \texorpdfstring{$\calQ$}{Q}}
In this section we provide some further details on the derivation of the boundary of $\calQ$, the set of all quantum states, when a few observables are known.
To derive these bounds, we need to find the conditions on the coefficients of Hermitian matrices with trace lesser than $1$, that ensure their being positive semidefinite.
More formally, the task is the following: given a set of observables $(\calO_1,...,\calO_n)$, what are the possible sets of expectation values $(\Tr(\calO_i \rho))_{i=1}^n$ when $\rho$ ranges across all possible states?
As a trivial example, if we consider the single observable $P_0\equiv\ketbra00$, then clearly the possible values it can take are $0\le \Tr(P_0\rho)\le1$, which we will write here concisely as just $0\le P_0\le 1$.
On the other hand, if we are given the single observable $X_{01}$, then its possible values are $-1\le X_{01}\le 1$, with extremal points achieved for the states $\frac1{\sqrt2}(\ket0\pm\ket1)$.
Note how these boundaries different from the corresponding boundaries of $\calC$: for example, $X_{01}=1$ is not possible with classical states.

\parTitle{Boundary in the space $(P_0,P_1,X_{01},Y_{01})$}
For example, if the observed quantities are $P_0,P_1, X_{01}, Y_{01}$, we need to study under what conditions we have
\begin{equation}
    \begin{pmatrix}
        P_0 & (X_{01}-iY_{01})/2 \\
        (X_{01}+iY_{01})/2 & P_1
    \end{pmatrix} \ge0,
\end{equation}
with the additional constraint $P_0+P_1\le1$, to ensure that this matrix can be obtained projecting a higher-dimensional (normalized) state.
From Sylvester's criterion~\cite{johnson1985matrix}, this is seen to be equivalent to the constraints $P_0, P_1\ge0$, and
\begin{equation}
    X_{01}^2+Y_{01}^2\le 4 P_0 P_1.
\end{equation}
The same results apply when the expectation values of $P_i,P_j,  X_{ij}$ and $Y_{ij}$ are known, for any $0\le i<j$.
If only a single coherence term $R_{ij}(\theta)$ is given, then the boundary corresponds to $|R_{ij}(\theta)|\le 2 \sqrt{P_i P_j}$.

\parTitle{Constraints with three coherences}
The constraints when three diagonal terms are known are instead more complicated. In this case, it is necessary to find the conditions such that
\begin{equation}\small
    \begin{pmatrix}
        P_0 & \dfrac{X_{01}-iY_{01}}{2} & \dfrac{X_{02}-iY_{02}}{2} \\
        \dfrac{X_{01}+iY_{01}}{2} & P_1 & \dfrac{X_{12}-iY_{12}}{2} \\
        \dfrac{X_{02}+iY_{02}}{2} & \dfrac{X_{12}+iY_{12}}{2} & P_2
    \end{pmatrix} \ge0.
\end{equation}
Let us temporarily denote with $c_{ij}$ the off-diagonal matrix elements.
From the determinants of the principal submatrices, the positive semidefiniteness condition is equivalent to the three conditions $|c_{ij}|\le \sqrt{P_i P_j}$, which reflect the same conditions we obtain when knowing only a coherence term and the corresponding probabilities, and an additional constraint of the form
\begin{equation}\small
    |t_{01}|^2 + |t_{02}|^2 + |t_{12}|^2 \le
    1 + 2\operatorname{Re}(t_{01}t_{12}\bar t_{13}),
\end{equation}
where $t_{ij}\equiv c_{ij}/\sqrt{P_i P_j}$.
This gives us a direct geometric understanding of the boundary of the space of all states. For every value of $P_0, P_1, P_2$, we get a corresponding range of possible values for off-diagonal elements. Similar results can be found, via careful application of Sylvester's criterion, for larger sets of observables.

\section{Toy example: \texorpdfstring{$(P_0,P_1)$}{(P0,P1)}}

In this section we show in a simple example how support functions can be used to derive boundaries. We focus on the space $(P_0,P_1)$, with $P_i\equiv \ketbra i$ projections onto the Fock basis states.
Our goal is showcasing, in a well-known simple case, what the derivation of quantum and classical boundaries via support functions would look like.

\parTitle{Quantum boundary}
Computing the boundary of $\calQ$ in the space $(P_0,P_1)$ is trivial, as we know that all pairs of non-negatives with $0\le P_0+P_1\le1$ are achievable with some quantum state.

\parTitle{Classical boundary}
The support function $h_\calC$ reads in this case, writing $\bsn\equiv(a,b)\in\mathbb{R}^2$,
\begin{equation}\label{eq:support_function_in_P0P1}
    \small
    h_\calC(a,b) =
    \max_{\rho\in\calC} (a P_0(\rho) + b P_1(\rho))
    =  \max_{\mu\ge0} e^{-\mu}(a + b \mu).
\end{equation}
Being the support function homogeneous, $h_\calC(\lambda a,\lambda b)=\lambda h_\calC(a,b)$ for all $\lambda\ge0$, we can restrict our attention to the cases with $b=1$ and $b=-1$. More specifically, the nontrivial boundary arises from the $b=1$ sector, so we will restrict to this case in the following.
Defining
$f(\mu;a)\equiv e^{-\mu} (a + \mu)$, we find that
\begin{equation}
    \partial_\mu f(\mu;a)=e^{-\mu}(1 - a - \mu).
\end{equation}
It follows that $\mu=1-a$ is a stationary point for $f$, which can also be verified to be a local maximum.
Thus the maximization in $h_\calC(a)\equiv h_\calC(a,1)$ can be achieved via either $\mu=1-a$, $\mu=\infty$, or $\mu=0$, and
$
    h_\calC(a,1) = \max(e^{a-1}, a, 0).
$
We can then observe that, for all $a\in\mathbb{R}$, we have $e^{a-1}\ge0$ and $e^{a-1}\ge a$, and thus conclude that 
\begin{equation}
    h_\calC(a,1)=e^{a-1}.
\end{equation}
This result provides a nonclassicality criterion for each value of $a$.
To otain a general nonclassicality criterion, which accounts for all the criteria with different $a$ at the same time, we observe that our goal is, given some observed values of $P_0$ and $P_1$, find $a$ such that
\begin{equation}
    a P_0 + P_1 > h_\calC(a,1).
\end{equation}
We can thus obtain a general nonclassicality criterion computing
\begin{equation}
    H_\calC(P_0)\equiv
    \min_a [h_\calC(a,1) - aP_0].
\end{equation}
Computing this amounts to finding the value of $a$ that is optimal to certify the nonclassicality of given $P_0$ and $P_1$.
Defining $H_\calC(a;P_0)\equiv h_\calC(a,1)-a P_0$, we see easily that $\lim_{a\to\pm\infty}H_\calC(a;P_0)=\infty$, which means that the minimum is achieved at a local stationary point. The function is continuous, and therefore we can find the minimum by imposing $\partial_a H_\calC(a;P_0)=0$, which gives
$
    a_{\rm min} = \log P_0 + 1,
$ 
from which we get
\begin{equation}
    H_\calC(P_0) = P_0 \log\left(\frac1 P_0\right),
\end{equation}
and conclude that the relevant nonclassicality criterion is $P_1 > P_0 \log(1/P_0)$.
Of course, this result could have been obtained by more direct methods, and was already discussed at length in~\cite{innocenti2020nonclassicality}, but it is nonetheless highly interesting to derive it like this, as this approach is much more suitable for computing criteria in more general spaces.

\section{Derivation of other two-dimensional criteria}

However, for classical states, there is in general no simple way to write a bound for $P_j$ in terms of $P_k$, albeit it is possible to express algebraically their relations via implicit relations.
For example, considering $(P_1,P_2)$, the set of coherent states traces the curve
$\frac{2 P_2}{P_1^2} \exp(-2P_2/P_1) =1$,
but there is no algebraic expression in terms of simple functions of the bound on $P_1$ for any given value of $P_2$.
Nonetheless, if one can find such relations \textit{e.g.} numerically, this provides bounds of the form $P_j^m(P_i) \le P_j \le P_j^M(P_i)$ satisfied by classical states, which then translates into an inequality in the $(X_{ij},P_i)$ space of the form
\begin{equation}
    0 \le |X_{ij} | \le 2\sqrt{P_i P_j^M(P_i)}.
\end{equation}
For example, if $P_1=0.2$, then $|X_{12}|\le 2\sqrt{0.2\times 0.254}\simeq 0.45<1$, and thus any measured value of $X_{12}>0.45$ certifies nonclassicality.
Alternatively, one can retrieve directly the algebraic relation between $X_{ij}$ and $P_i$ satisfied by all coherent states using ideas similar to those discussed in~\cite{innocenti2020nonclassicality} for Fock states.
More specifically, one can always find coefficients $\alpha,\beta,\gamma\in\RR$, depending on $i,j\in\mathbb N$, such that
\begin{equation}
\begin{aligned}
    X_{ij}^2/P_i^\alpha = e^{-\beta \mu},
    \qquad
    \mu = (X_{ij}/P_i)^\gamma,
\end{aligned}
\end{equation}
and thus the set of classical states is in the convex hull of the curve defined algebraically as
\begin{equation}
    \frac{X_{ij}^2 }{P_i^\alpha} \exp\left[-\beta\left(\frac{X_{ij}}{P_i}\right)^\gamma\right] = 1.
\end{equation}


\section{Example with coherences providing enhanced predictive power}

As hinted at in the main text, there are situations where knowledge of coherences allows to detect nonclassicality that would not have been otherwise detectable via only Fock-state probabilities.

We consider here as such an example the space $(P_0,P_2,X_{02})$.
We know that the set of coherent states in non-convex in the space $(P_0,P_2)$, as can be seen directly from~\cref{fig:nonclassicality_attenuated0p2_mergedVertically}. Let us consider a point in this non-convex region, for example, $P_0=0.6$ and $P_2=0.1$.
This is thus clearly classical given only knowledge of $(P_0,P_2)$. Is it possible to detect it as nonclassical if knowledge of $X_{02}$ is provided?
To see that this is the case, we can consider the associated support function, and show that there are values of $(a,b,c)\in\mathbb{R}^2$ such that
\begin{equation}
    a X_{02} + b P_0 + c P_2 > h_\calC(a,b,c),
\end{equation}
where
\begin{equation}
\begin{gathered}
    h_\calC(a,b,c)\equiv\sup_{\mu\ge0,\varphi}
    e^{-\mu}\left[
        \sqrt2 a\mu\cos(2\varphi)+b +c \frac{\mu^2}2
    \right] \\
    = \sup_{\mu\ge0} e^{-\mu} \left[
    |a| \sqrt2 \,\mu+b+c\frac{\mu^2}{2}
    \right].
\end{gathered}
\end{equation}
Let us consider the case with $a=1/2$ and $c=1$.
With this assumption the calculation of the support function simplifies to
\begin{equation}\label{SM:eq:support_function_special_values_ac}
    h_\calC(0.5,b,1) = \frac12 \sup_{\mu\ge0} e^{-\mu} \left[
    \mu^2 + \sqrt2 \mu + 2b \right].
\end{equation}
The associated function to maximize is $f(\mu;b)\equiv\mu^2+\sqrt2\mu+2b$, which can be seen to have local real stationary points, for $b\le3/4$, at
\begin{equation}
    \mu_\pm(b) = \frac{2-\sqrt2}{2} \pm \frac{\sqrt{6-8b}}{2},
\end{equation}
with $\mu_-(b)\ge0$ only for $b\ge1/\sqrt2$.
One can further verify that $f(\mu_+(b);b)\ge f(\mu_-(b);b)$, meaning that only the $\mu_+(b)$ local solution needs be considered. We also have $f(\mu_+(b);b)\ge0$ in the domain of definition of $\mu_+$, which allows us to exclude the $\mu\to\infty$ solution to~\cref{SM:eq:support_function_special_values_ac}. Finally, $f(0;b)=b$, which for $b>0$ is a possible alternative solution.
The transition between the local solution and the $\mu=0$ solution happens when
\begin{equation}
    f(\mu_+(b);b) = f(0;b) = b.
\end{equation}
This equation has no analytic solution, but can be verified numerically to pinpoint $b_{\rm tr}\simeq 0.738$.

\begin{figure}
    \centering
    \includegraphics[width=\linewidth]{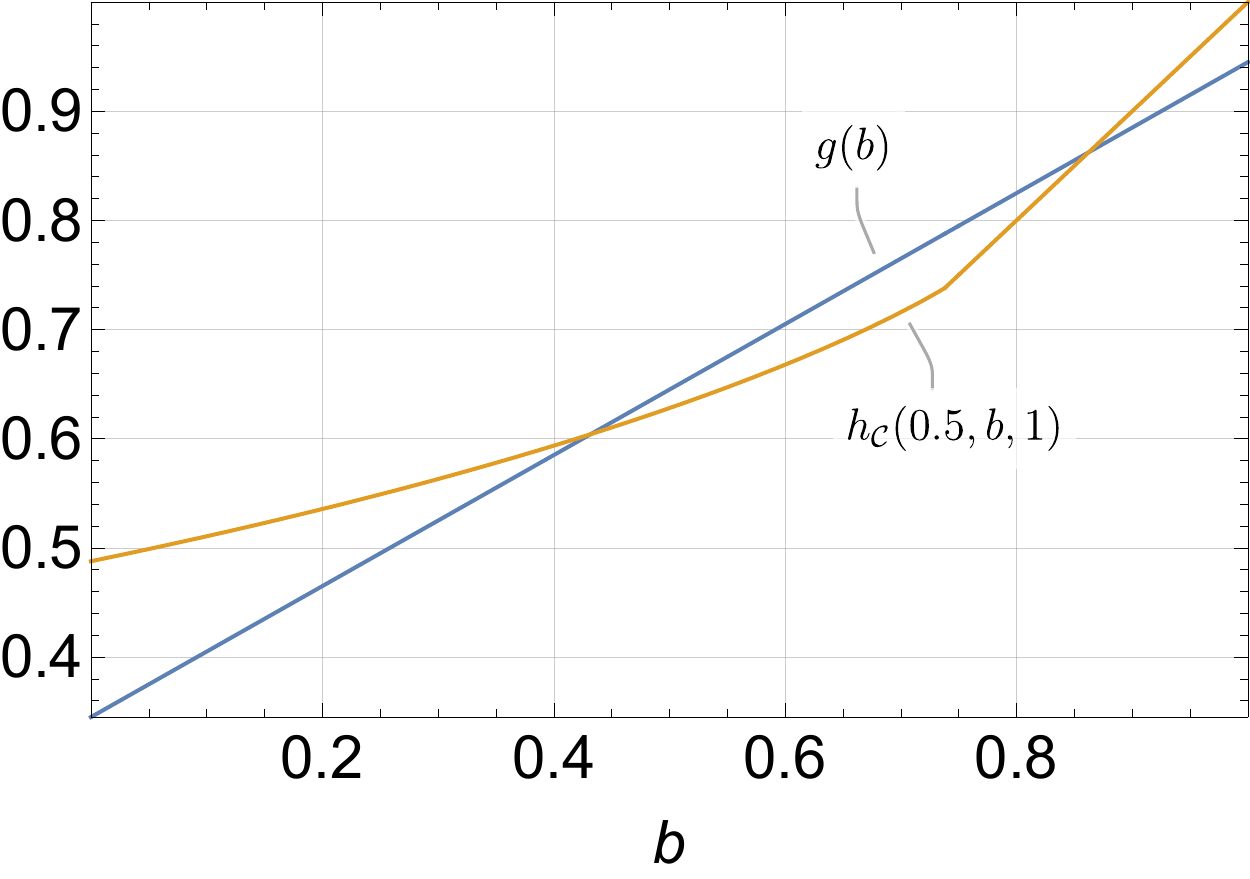}
    \caption{Support function $h_\calC(0.5,b,1)$ and $g(b)$ as a function of $b$. The values of $b$ corresponding to $g(b)>h_\calC(0.5,b,1)$ can all be used to certify the nonclassicality of the corresponding observed quantities, here $(P_0,P_2,X_{02})$ with $P_0=0.6$, $P_2=0.1$, and $X_{02}=2\sqrt{P_0P_2}$.}
    \label{fig:suppFunctionP0P2X02_specialcase}
\end{figure}

All of the above reasoning allows us to finally write down the support function as
\begin{equation}
\begin{cases}
    h_\calC(0.5,b,1) = f(\mu_+(b);b), & b \le b_{\rm tr}, \\
    h_\calC(0.5,b,1) = b, & b>b_{\rm tr}.
\end{cases}
\end{equation}
With this explicit expression for the support function we can answer probe for the existence of some $b$ such that
\begin{equation}
    0.5 X_{02} + b P_0 + P_2 > h_\calC(0.5,b,1),
\end{equation}
for some $X_{02}$. More specifically, we can consider a case where the coherence term is saturated, meaning $X_{02}=2\sqrt{P_0 P_2} \simeq 0.49$.
All that remains is now to compare the function
$g(b)\equiv 0.5 X_{02} + b P_0 + P_2$, for the given values of $P_0,P_2,X_{02}$, and see whether there are values of $b$ such that $g(b)>h_\calC(0.5,b,1)$.
As can be seen clearly from~\cref{fig:suppFunctionP0P2X02_specialcase}, any $0.6\le b\le 0.8$ will work to certify nonclassicality.

We conclude that there are situations where coherence terms are necessary to detect the nonclassicality. As a more explicit examples, a state resulting in the observed values considered above is
\begin{equation}
    \rho\equiv 0.7 \ketbra\psi + 0.3 \ketbra3,
\end{equation}
with $\ket\psi\equiv \frac{1}{\sqrt{0.7}} (\sqrt{0.6}\ket0 + \sqrt{0.1}\ket2)$.
This state gives $P_0=0.6, P_2=0.1$, $X_{02}=2\sqrt{P_0 P_2}$, cannot be detected as nonclassical looking only at $P_0$ and $P_2$, but can be detected as nonclassical using the coherence term $X_{02}$, exploiting the inequality obtained from the support function with values $a=0.5,b=0.6, c=1$.

\section{Effects of attenuation}

In this section we provide, for completeness, explicit expressions for the states obtained via attenuation from superpositions of Fock states, and show other examples of how their nonclassicality emerges from coherence terms.

\subsection{\texorpdfstring{$\frac{1}{\sqrt2}(\ket0+\ket1)$}{0+1}}
Consider the superposition of Fock states
$\ket\psi\equiv \frac{1}{\sqrt2}(\ket0+\ket1)$.
Evolving through a beamsplitter with transmissivity $t\in\CC$. The output state on the first output mode is
\begin{equation}
    \rho_t = \frac12[(2-T) P_0 + T P_1 + (t\ketbra{1}{0}+\on{h.c.})],
\end{equation}
where $T\equiv\abs{t}^2$ is the beamsplitter transmissivity.
The corresponding expectation values read
\begin{equation}
    \expval{P_0} = \frac{2-T}{2},
    \quad \expval{P_1} = \frac{T}{2},
    \quad \expval{X_{01}} = \sqrt{T}\cos(\phi),
\end{equation}
where $t=\sqrt{T}e^{i\phi}$.

\subsection{\texorpdfstring{$\frac{1}{\sqrt2}(\ket0+\ket2)$}{0+2}}
\begin{figure}
    \centering
    \includegraphics[width=\linewidth]{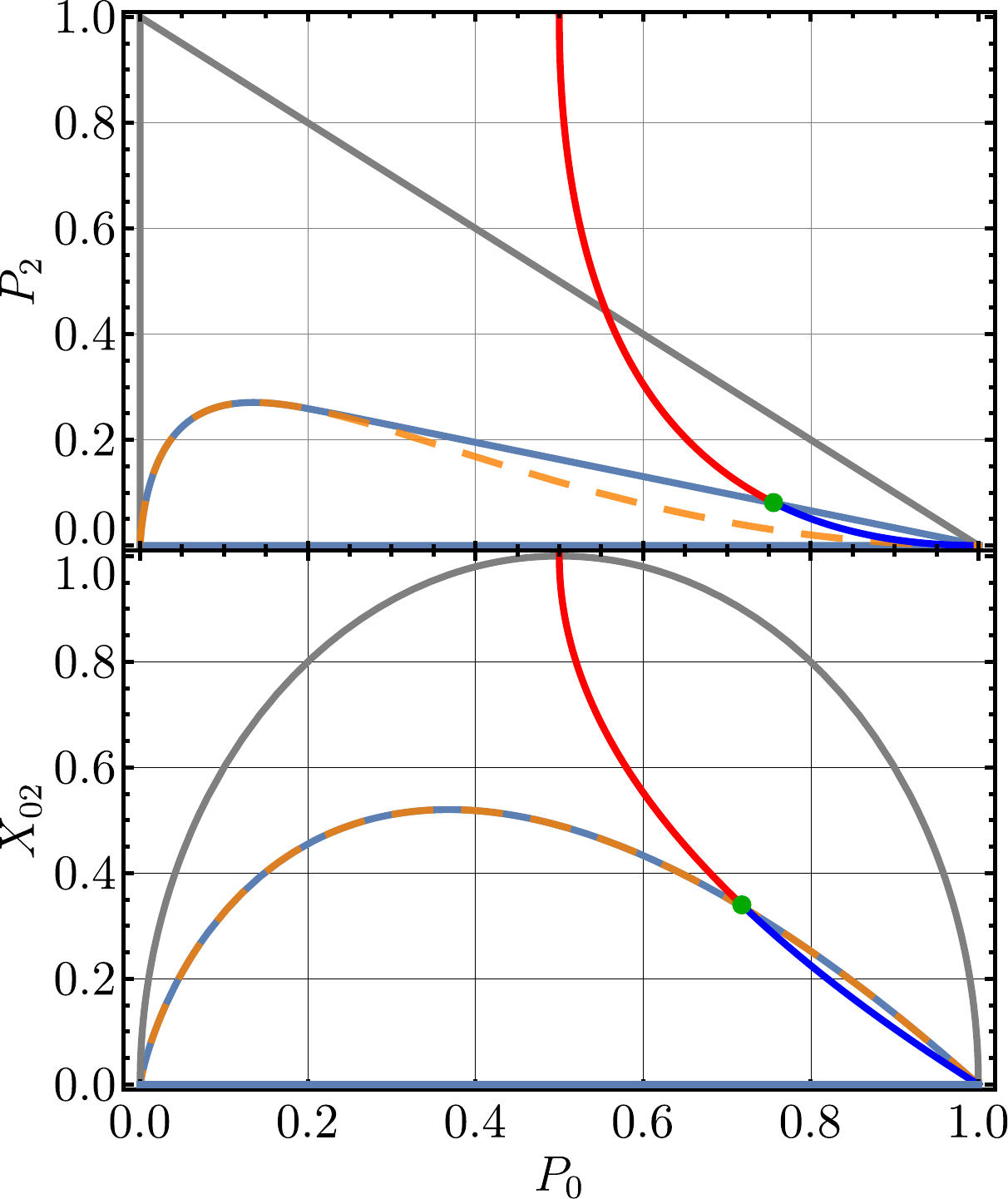}
    \caption{States obtained attenuating $\frac{1}{\sqrt2}(\ket0+\ket2)$ with different transmissivities, in the spaces $(P_0, P_2)$ (\textbf{\textit{a}}) and $(P_0, X_{02})$ (\textbf{\textit{b}}). The blue filled regions are bounded by the coherent states in the respective spaces. The classical region is thus the \textit{convex hull} of these regions. While in (\textit{\textbf{b}}) the boundary is convex and thus the classical set coincides with the blue filled region, the same is not true in (\textit{\textbf{a}}). The purple dotted line gives the values of the support function $h_{\calC}(u)$ for all the possible directions $u\in S^1$. This is what is used to determine classicality of a state (equivalently, it is used to determine the convex hull of the coherent states). The red and blue continuous line corresponds to the states obtained attenuating $\frac{1}{\sqrt2}(\ket0+\ket2)$. The red sections correspond to nonclassical states. The green dot between the two marks the transition between classicality and nonoclassicality.}
    \label{fig:nonclassicality_attenuated0p2_mergedVertically}
\end{figure}

If the input state is a superposition of vacuum and Fock state $\ket2$, after evolution through the beamsplitter we get
\begin{equation}
\begin{gathered}
    \rho_t =
    \frac12 (1+R^2)\PP_0
    + TR \PP_1
    + \frac12 T^2 \PP_2 \\
    + \frac12 (t^2\ketbra{2}{0}+\bar t^2\ketbra{0}{2}),
\end{gathered}
\end{equation}
where $R\equiv 1-T$. The corresponding expressions for the expectation values of $P_0, P_1, P_2, X_{02}$ follow directly from this.
The nonclassicality of these attenuated states in the $N=2$ case is shown in~\cref{fig:nonclassicality_attenuated0p2_mergedVertically}.
In particular, we find nonclassicality in $(P_0, X_{02})$ for $T\ge 0.34$, and in $(P_0, P_2, X_{02})$ for $T\ge 0.273$.

\subsection{\texorpdfstring{$\frac{1}{\sqrt2}(\ket1+\ket2)$}{1+2}}
If we consider the input state $\frac{1}{\sqrt2}(\ket1+\ket2)$, the nonvanishing coherence terms after attenuation through a beamsplitter are $X_{01}$ and $X_{12}$. We find the nonclassicality threshold to correspond to the attenuation $T\approx0.84$, as shown in~\cref{fig:convex_hulls_X01vs} in the main text. As in the previous cases, we therefore find a nontrivial nonclassicality threshold for the coherences.

\section{Effects of thermal noise}
\label{app:thermal_noise_equations}

We outline here for completeness the procedure used to obtain the results about thermalized attenuated states, reported in the main text.

Thermal noise can be modelled as the following operation applied to a given state $\rho_t$:
\begin{equation}\scalebox{0.9}{$\displaystyle
    \Phi^{\rm th}_{\bar n}(\rho_t)
    \equiv \int d\mu \frac{e^{-\mu/\bar n}}{\bar n}
    \int_0^{2\pi} \frac{d\phi}{2\pi}
    D(\sqrt\mu e^{i\phi})\rho_t D^\dagger(\sqrt\mu e^{i\phi}),
$}\end{equation}
where $D(\alpha)=\exp(\alpha a^\dagger-\bar\alpha a)$ is the displacement operator.



When $\rho_t$ is obtained evolving $\frac{1}{\sqrt2}(\ket0+\ket1)$ through a beamsplitter with transmittivity $t$, we find that
\begin{equation}
\begin{gathered}
	\Phi_{\bar n}^{\rm th}(\rho_t)
	= \sum_{j=0}^\infty \left(
	P_j^{(\mathrm{th};t, \bar n)}
	\PP_j
	+ X_j^{(\mathrm{th}; t,\bar n)} X_j
	\right),
\end{gathered}
\end{equation}
where $X_j \equiv \ketbra{j}{j+1} + \ketbra{j+1}{j}$, and
\begin{equation}
\begin{aligned}
    P_j^{(\mathrm{th};t, \bar n)}
    &\equiv \frac{\bar n^j}{(\bar n + 1)^{j+1} }
    \frac{[2\bar n(\bar n + 1) + T(j - \bar n)]}{2\bar n (\bar n + 1)}, \\
    X_j^{(\mathrm{th}; t,\bar n)}
    &\equiv
    \frac{\bar n^j}{(\bar n + 1)^{j+2}} \sqrt{j+1}
    \Re(t).
\end{aligned}
\end{equation}
In particular,
\begin{equation}
\begin{aligned}
    P_0^{(\mathrm{th};t, \bar n)}
    &= \frac{1}{2\bar n (\bar n + 1)^{2} }
    [2\bar n(\bar n + 1) - T \bar n], \\
    P_1^{(\mathrm{th};t, \bar n)}
    &= \frac{1}{2 (\bar n + 1)^{3} }
    [2\bar n(\bar n + 1) + T(1 - \bar n)], \\
    P_2^{(\mathrm{th};t, \bar n)}
    &= \frac{\bar n^2}{2\bar n (\bar n + 1)^{4} }
    [2\bar n(\bar n + 1) + T(2 - \bar n)], \\
    X_0^{(\mathrm{th};t,\bar n)}
    &= \frac{\Re(t)}{(\bar n + 1)^2}.
\end{aligned} 
\end{equation}

Similar calculations are used to derive the expressions in the case of $\ket0+\ket2$, used for Fig. 3 in the main text.

\end{appendices}

\bibliography{main}
\bibliographystyle{apsrev4-2}

\end{document}